\documentclass{article}

\usepackage{microtype}
\usepackage{graphicx}
\usepackage{subfigure}
\usepackage{booktabs} 
\usepackage{multirow}
\usepackage{gensymb}
\usepackage{siunitx}
\usepackage{caption}
\usepackage[dvipsnames]{xcolor}

\usepackage[most]{tcolorbox}

\newtcolorbox{examplebox}{
  colback=blue!5!white,
  colframe=blue!75!black,
  fonttitle=\bfseries, 
  boxrule=0.8pt,
  arc=4pt,
  top=6pt, bottom=6pt,
  left=6pt, right=6pt
}

\usepackage{hyperref}

\usepackage[accepted]{icml2025}

\usepackage{amsmath}
\usepackage{amssymb}
\usepackage{mathtools}
\usepackage{amsthm}

\usepackage[capitalize,noabbrev]{cleveref}

\theoremstyle{plain}

\theoremstyle{definition}

\theoremstyle{remark}

\usepackage[textsize=tiny]{todonotes}

\definecolor{MyGold}{HTML}{FFD700}    
\definecolor{MySilver}{HTML}{C0C0C0}  
\definecolor{MyBronze}{HTML}{CD7F32}  
\definecolor{MyFourth}{HTML}{8FBC8F}  

\icmltitlerunning{Generative Port of Mars}

\begin{document}

\twocolumn[
\icmltitle{Using Large Language Models to Simulate Human Behavioural Experiments: Port of Mars}



\icmlsetsymbol{equal}{*}

\begin{icmlauthorlist}
\icmlauthor{Oliver Slumbers}{ucl}
\icmlauthor{Joel Z. Leibo}{gdm}
\icmlauthor{Marco A. Janssen}{asu}
\end{icmlauthorlist}

\icmlaffiliation{ucl}{University College London}
\icmlaffiliation{gdm}{Google DeepMind}
\icmlaffiliation{asu}{Arizona State University}

\icmlcorrespondingauthor{Oliver Slumbers}{oliver.slumbers.19@ucl.ac.uk}

\icmlkeywords{Machine Learning, ICML}

\vskip 0.3in
]



\printAffiliationsAndNotice{} 

\begin{abstract}
Collective risk social dilemmas (CRSD) highlight a trade-off between individual preferences and the need for all to contribute toward achieving a group objective. Problems such as climate change are in this category, and so it is critical to understand their social underpinnings. However, rigorous CRSD methodology often demands large-scale human experiments but it is difficult to guarantee sufficient power and heterogeneity over socio-demographic factors. Generative AI offers a potential complementary approach to address this problem. By replacing human participants with large language models (LLM), it allows for a scalable empirical framework. This paper focuses on the validity of this approach and whether it is feasible to represent a large-scale human-like experiment with sufficient diversity using LLM. In particular, where previous literature has focused on political surveys, virtual towns and classical game-theoretic examples, we focus on a complex CRSD used in the institutional economics and sustainability literature known as Port of Mars.
\end{abstract}

\section{Introduction}
\label{introduction}
\begin{figure*}[ht!]
\centering
\includegraphics[width=1\linewidth]{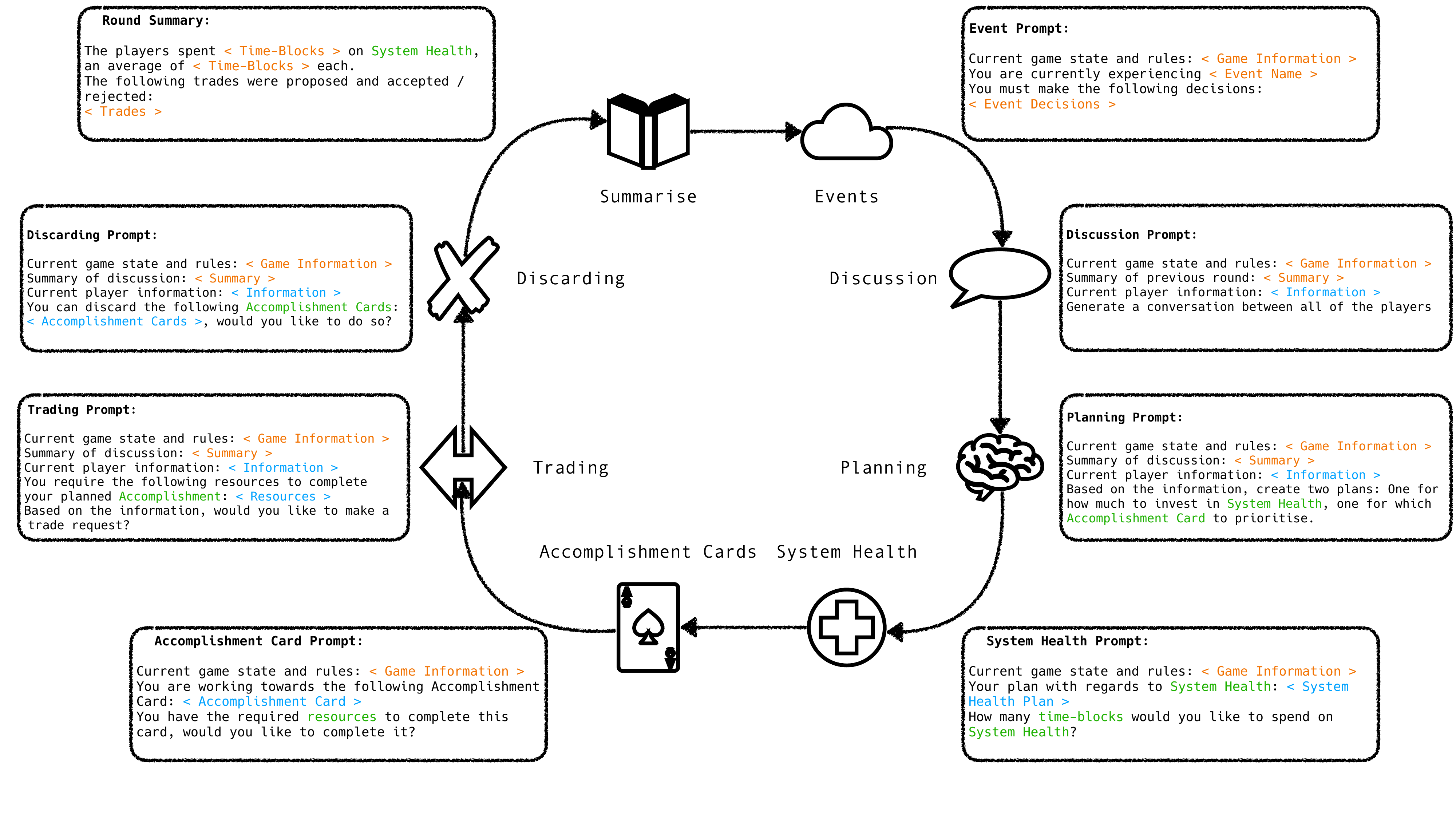}
\caption{Visualisation of a single round. We show the general format of the prompts that are passed to the players, these are simplified and detailed fully in Appendix \ref{appendix:prompts}. Items in \textcolor{ForestGreen}{green} refer to Port of Mars specific terminology. Items in \textcolor{orange}{orange} refer to public information. Items in \textcolor{blue}{blue} refer to private player information.}
\label{fig:pom-flow}
\end{figure*}

A Collective Risk Social Dilemma (CRSD) is a situation where a group must reach a collective target through individual contributions where everybody suffers if they fail to achieve the target \cite{milinski2021collective}. CRSD encompass many of the challenges facing society today. Some examples include emission reductions to reduce risks of climate change, and the vaccination for infectious diseases to create herd immunity.

Societal challenges related to the need for collective action are well-studied, in particular in trying to understand the governance of shared resources (e.g. land, forests, water, etc...) \cite{hardin1968thetragedyofthecommons, ostrom2008tragedy}. \citet{hardin1968thetragedyofthecommons} introduced the term `tragedy of the commons', encapsulating the notion that communities fail to govern shared resources without a central government regulating them, or private property systems. \citet{ostrom1992community} later demonstrated that communities can self-govern under certain conditions. Still, more understanding is needed about the conditions that encourage collective action and this research span multiple disciplines \cite{schill2019more}.

Experimental social science, rather than observational social science, provides us with a method to approach this question \cite{falk2009lab}. Artificial experiments in lab settings provide controlled variation, for example helping social scientists rule out confounding effects when analysing the factors contributing to collective action. Ascertaining causality can prove difficult with the kinds of data available to observational social scientists \citep{morgan2007counterfactuals}. This is why experimental CRSD research is needed. However, this is afflicted by low power due to the difficulty, and cost, of large-scale human experiments. It is difficult to create large enough samples to validate results, and 
insufficient heterogeneity amongst participants can limit generalizability. Our hypothesis, along with others \cite{dillion2023can, xu2024ai}, is that large language models (LLMs) can be used to mitigate some obstacles that would otherwise make experimental CRSD research difficult. LLMs will not replace experimental research but can provide a useful complement, for example by identifying relevant configurations of experiments with human participants to implement and to provide predictions for those experiments.

LLMs have the potential to model knowledge, beliefs and patterns of social behaviour, facilitating \textit{in silico} research on human social behaviour \cite{argyle2023out, aher2023using, grossmann2023ai}. Due to being trained on a large corpus of human data, it is plausible to expect that LLMs \textit{may} contain substantial latent social information \cite{horton2023large}. LLMs could be cost-effective / efficient tools to aid empirical studies by narrowing the pool of hypotheses and gaining guiding insights. However, the conditions that must hold for LLM-``participants'' to be viable as a tool of \textit{in silico} research remain an equally open-question.

\citet{argyle2023out} introduced \textit{Algorithmic Fidelity}: `the degree to which the complex patterns of relationships between ideas, attitudes, and sociocultural contexts within a model accurately mirror those within a range of human subpopulations'. The main idea is that LLM responses are coherent with those of distinct human subpopulations can be elicited by prompting the model to continue a response from the perspective of a member of the group in question. They suggested four minimum conditions to check for the validity of a given piece of LLM-based \textit{in silico} research:

\begin{enumerate}
\item \textbf{Social Science Turing Test}---LLM responses are indistinguishable from parallel human texts.
\item \textbf{Backward Continuity}---LLM responses are consistent with the attitudes and sociodemographic information of the conditioning context, such that humans viewing responses can infer key elements of that input.
\item \textbf{Forward Continuity}---LLM responses proceed naturally from the conditioning context provided, reliably inflecting the form, tone and content of the context.
\item \textbf{Pattern Correspondence}---LLM responses reflect underlying patterns of relationships between ideas, demographics, and behaviour that would be observed in comparable human-produced data.
\end{enumerate}

When all four criteria hold then we can say there is algorithmic fidelity, and the implication is that it would be valid to conduct \textit{in silico} research on LLM agents in such a setting\footnote{There are a great many caveats, and the full methodological significance of algorithmic fidelity is far from settled. For further discussion see \cite{argyle2023out, amirova2024framework}.}.

Algorithmic fidelity has predominantly received interest in domains where the goal is to characterize a distribution of static human responses (e.g.~survey data \citet{argyle2023out} and structured interview data \citet{amirova2024framework}). These studies did not put multiple LLM agents together in the same environment to study their social interactions or how they solve problems together. Seperately, another line of work has focused on socially interacting LLM agents. Generative agents, such as those studied in \cite{park2023generative}, have been used to simulate groups of interacting individuals. These methods have potential in agent-based modeling of collective action dilemmas \citep{poteete2010working}. They enable the creation of a new kind of agent-based model where it is possible to study the emergence of group-level social phenomena from the interaction of individual agents. These individuals can now talk to each other and utilize much more complex kinds of cognition than was possible with simpler agents \citep{vezhnevets2023generative, ghaffarzadegan2024generative}.

In this work, we explore LLM agents in a novel CRSD. We initially validate the algorithmic fidelity of LLM agent responses within the CRSD by directly comparing against human responses and metrics from the original works. Then, based on an underlying assumption of algorithmic fidelity, we demonstrate the flexibility and power to model and study the interactions of differing social personas within CRSDs. In particular, we focus on Social Value Orientation (SVO) as the primary way of defining differences between our agents and study how this impacts the outcomes in the CRSD. Furthermore, we study the effects of leadership on top of the SVO outcomes, demonstrating the flexibility of traits and problem variants that can be effectively studied under an LLM agent approach. 

\section{Port of Mars}

 Collective action problems are widely studied \cite{ostrom2000collective}, but they are generally abstracted as simple matrix games \cite{trivers1971evolution, axelrod1981evolution}. While such abstractions have been used for decades, many have argued that they are of limited utility for predicting human behaviour in more complex (and realistic) settings \cite{janssen2010lab, perolat2017multi, hertz2024beyond}. 
In particular, matrix games generally ignore critical features of real-world social dilemmas \cite{leibo2017multi}. Firstly, they miss the temporal extendedness of real-world dilemmas. Secondly, they do not require decisions with only partial information about the state of the world, the activities of others and true uncertainty. For example, the true impact of the climate crisis and human action towards it is an unknown, and unexpected events will occur regularly. In traditional public goods / common pool resources experiments, deep uncertainty is often elided since the probabilities of events are known (e.g. the probabilities of stochastic events in a poker game). Therefore, it is important to ensure any new framework is capable of introducing  `unknown unknowns' to represent the ``sometimes we just have to cross that bridge when we come to it'' feature of real-world dilemmas \citep{knight1921risk, savage1951foundations, binmore2007rational, johanson2022emergent}.
 

We must also consider how LLMs may impact methodological decisions, such as
player motivation. Often monetary rewards are attached to performance, however, it is debated whether this an effective motivator \cite{smith1993monetary}. 
\citet{gneezy2000pay} found that appealing to moral incentives may be more effective. 
In contrast to AI research methods, like reinforcement learning, LLMs have no natural analog of reward. Therefore it is unnatural to motivate LLMs with monetary incentives. Instead, LLM agents can be motivated in other ways, e.g.~by appealing to their social identity, or by asking them to role play as a certain character. We must choose a CRSD that appeals to such incentives in order to motivate behaviour. However, a critical problem is the extensive literature that exists about CRSD. This contaminates the LLM training data, making it difficult to separate the LLMs social capabilities from simply recreating strategies discussed in the literature. We require a near novel CRSD that is not prominent in the LLM training corpus.

 We have chosen the Port of Mars (PoM) \cite{janssen2020keeping} game for the following reasons: 1) PoM meets the criteria of \citet{hughes2018inequity} and thus is a \textit{sequential social dilemma}\footnote{See \citet{du2023review} for a review of these definitions.}, this may render it full of potential for emergent social complexity according to conjectures such as those of \citet{leibo2019autocurricula} 2) PoM focuses on building a narrative for the players to engage with. This naturally fits the LLM framework. 3) PoM is \textit{specifically} designed to include `unknown unknowns' (it is a game of incomplete information in the game-theoretic sense), 
 4) PoM is a relatively novel game. While it has been previously described \cite{janssen2020keeping}, it has a limited presence in the wider literature. This  suggests that training data contamination is unlikely to prove overly problematic.

PoM \cite{janssen2020keeping} is a resource allocation game where five players balance individual goals against the conflicting needs of maintaining shared infrastructure called System Health, in the face of ongoing environmental, social, and technical challenges. The narrative identifies the players as members of Generation Zero: the first group of long-term residents to arrive on the Red Planet, and the Players experience the challenges of life as early citizens of a Martian settlement. 
Full game details are provided in Appendix \ref{appendix:pom}.

Each player has a unique role: the Curator, Pioneer, Researcher, Politician, and Entrepreneur. Although the roles have a different narrative, the underlying payoff structure is the same. This is not known by the players. Shared infrastructure degrades due to wear and tear. Players must maintain a minimum level of shared infrastructure for the game to continue. Players spend personal resources on shared infrastructure or on acquiring points to win. 'Dirty cards' exist which players can claim for points at the expense of the shared infrastructure. As such, the game experience provides a fundamental CRSD between individual and collective goals. Each round features random event cards. These can lead to reductions of shared infrastructure (e.g. crop failures) or other positive / negative consequences. Normally, players draw one event card per round, however this increases as the shared infrastructure degrades. These events are a source of incomplete information, as the deck of event cards is unknown, and the LLM players do not retain memory of previous games to memorise event cards.


\section{Framework Overview}
All of the players in our framework are controlled by a single LLM, and they are differentiated through their prompting. All of the prompts, at minimum, contain the following information:

\begin{enumerate}
    \item \textit{Game Description} - A minimal description of the rules and narrative such that the LLM can play the game fully without requiring additional game information.
    \item \textit{Round Information} - Description of event cards that have been drawn in this round. Information about level of system health.
    \item \textit{Player Identity} - Provides the player role (e.g. Pioneer), and a description of their personality / traits.
    \item \textit{Phase Information} - Description of what currently needs to be done by the LLM player.
\end{enumerate}

Full prompts for each phase are provided in Appendix \ref{appendix:prompts}. Our version of PoM consists of 9 rounds, however the LLM agents are not aware of the number of rounds. In each round the players make the following decisions (also demonstrated in Fig. \ref{fig:pom-flow}):
\begin{enumerate}
    \item Handle specific decisions relating to events.
    \item Have a planning meeting with all of the other players.
    \item Create or update longer-term plans.
    \item Spend time-blocks on either the system health or the resources.
    \item Propose and accept/reject trade requests with other players for resources.
    \item Decide whether to complete / discard available accomplishment cards.
\end{enumerate}


An early problem we encountered was that providing full histories of the game did not lead to strong gameplay (e.g. forgetting the rules). However, without full history we often found that round-by-round behaviour was inconsistent, for example with players acquiring different resources across round that were not consistent with the accomplishment card they were aiming for in the earlier rounds. To overcome this, we carry three pieces of information through the rounds. First, we construct a summary of the round for the players that tracks information on trades that happened, system health spending in the previous the round, and an LLM generated summary of the planning meeting. Secondly, in order to generate more consistent decision-making, we get the LLM players to generate two plans that are used to condition future behaviour: 1) \textit{Health Plan} - The players decide how much of their time-blocks they want to invest in the system health, with their planning being contingent on the current rounds events and the player planning meeting, alongside the general information, and 2) \textit{Accomplishment Plan} - The players decide which of their available accomplishment cards they want to complete next. At the start of each round, players can either decide to continue working towards the accomplishment card they worked towards in the previous round, or update to a new one. We found empirically that without these elements it was difficult to get strong, consistent behaviour from the LLM players.



\section{Experiments}
\begin{table*}[!ht]
    \small
    \centering
    \begin{tabular}{l cc cc cc}
        \toprule
        \multirow{2}{*}{Personality Group}
         & \multicolumn{2}{c}{Points}
         & \multicolumn{2}{c}{\% Dirty Cards}
         & \multicolumn{2}{c}{Health Spend} \\
        \cmidrule(lr){2-3} \cmidrule(lr){4-5} \cmidrule(lr){6-7}
         & LLM Score & Human Rank
         & LLM Score & Human Rank
         & LLM Score & Human Rank \\
        \midrule
        \multicolumn{7}{l}{\textit{Forward Continuity}} \\
        \textbf{Selfish}
           & $8.8 \pm 0.8$ & N/A
           & 25\% & N/A
           & $8.3 \pm 1.9$ & N/A \\
        \textbf{Cooperative}
           & $3.7 \pm 0.4$ & N/A
           & 11\% & N/A
           & $40.8 \pm 1.9$ & N/A \\
        \midrule
        \multicolumn{7}{l}{\textit{Pattern Correspondence}} \\
        \textbf{Egalitarian Individualist}
           & \textcolor{MyGold}{$7.2 \pm 0.6$} & \textcolor{MyGold}{1 (2.7)} 
           & \textcolor{MySilver}{13\%} & \textcolor{MySilver}{2 (14\%)}             
           & \textcolor{MyBronze}{$17 \pm 1.7$} & \textcolor{MyGold}{$52.5 \ \pm 8.5$} \\ 
        \textbf{Egalitarian Communitarian}
           & \textcolor{MyBronze}{$5.0 \pm 0.3$} & \textcolor{MyFourth}{4 (3.7)} 
           & \textcolor{MyFourth}{4\%} & \textcolor{MyFourth}{4  (8\%)}               
           & \textcolor{MySilver}{$32 \pm 2.1$} & \textcolor{MyFourth}{$47.5 \pm 8.4$} \\ 
        \textbf{Hierarchical Individualist}
           & \textcolor{MySilver}{$6.2 \pm 0.5$} & \textcolor{MySilver}{2 (3.2)} 
           & \textcolor{MyGold}{15\%} & \textcolor{MyGold}{1 (14.3\%)}                 
           & \textcolor{MyFourth}{$13 \pm 1.4$} & \textcolor{MyBronze}{$51.4 \pm 4.2$} \\ 
        \textbf{Hierarchical Communitarian}
           & \textcolor{MyFourth}{$3.7 \pm 0.7$} & \textcolor{MySilver}{2 (3.2)} 
           & \textcolor{MyBronze}{7\%} & \textcolor{MyBronze}{3 (11.5\%)}               
           & \textcolor{MyGold}{$34 \pm 2.2$} & \textcolor{MySilver}{$51.7 \pm 7.3$} \\ 
        \bottomrule
    \end{tabular}
    \caption{Pattern Correspondence \& Forward Continuity results. The table is divided into two main sections indicated by sub-titles: 'Forward Continuity' (top two rows) and 'Pattern Correspondence' (bottom four rows).
    Within the 'Pattern Correspondence' section, colours indicate relative ranking for each metric:
    \textcolor{MyGold}{Gold} denotes 1st,
    \textcolor{MySilver}{Silver} denotes 2nd,
    \textcolor{MyBronze}{Bronze} denotes 3rd, and
    \textcolor{MyFourth}{Green} denotes 4th.
    In the Human Rank column we provide the corresponding personality group rankings from the human experiments established in \cite{janssen2020keeping}. Values in brackets for the Human Rank column represent the average finishing position (for the Points column) and raw percentage value (\% Dirty Cards column); data for Health Spend was not available in the source experiment. The Hierarchical Individualist and Communitarian received the same ranking in the points column in human experiments.}
  \label{table:new-main-table} 
\end{table*}

All experiments used gemini-1.5-flash-002 \cite{team2024gemini} for the LLM players. The purpose of this work is not to test the differences between specific LLMs (e.g. OpenAI vs. Gemini vs. Claude vs. etc...) behaviour in these social science experiments, and due to lack of resources we chose the gemini model as it provided free-tiered API requests.

\subsection{Algorithmic Fidelity}
\begin{figure*}[ht!]
\centering
\includegraphics[width=1\linewidth]{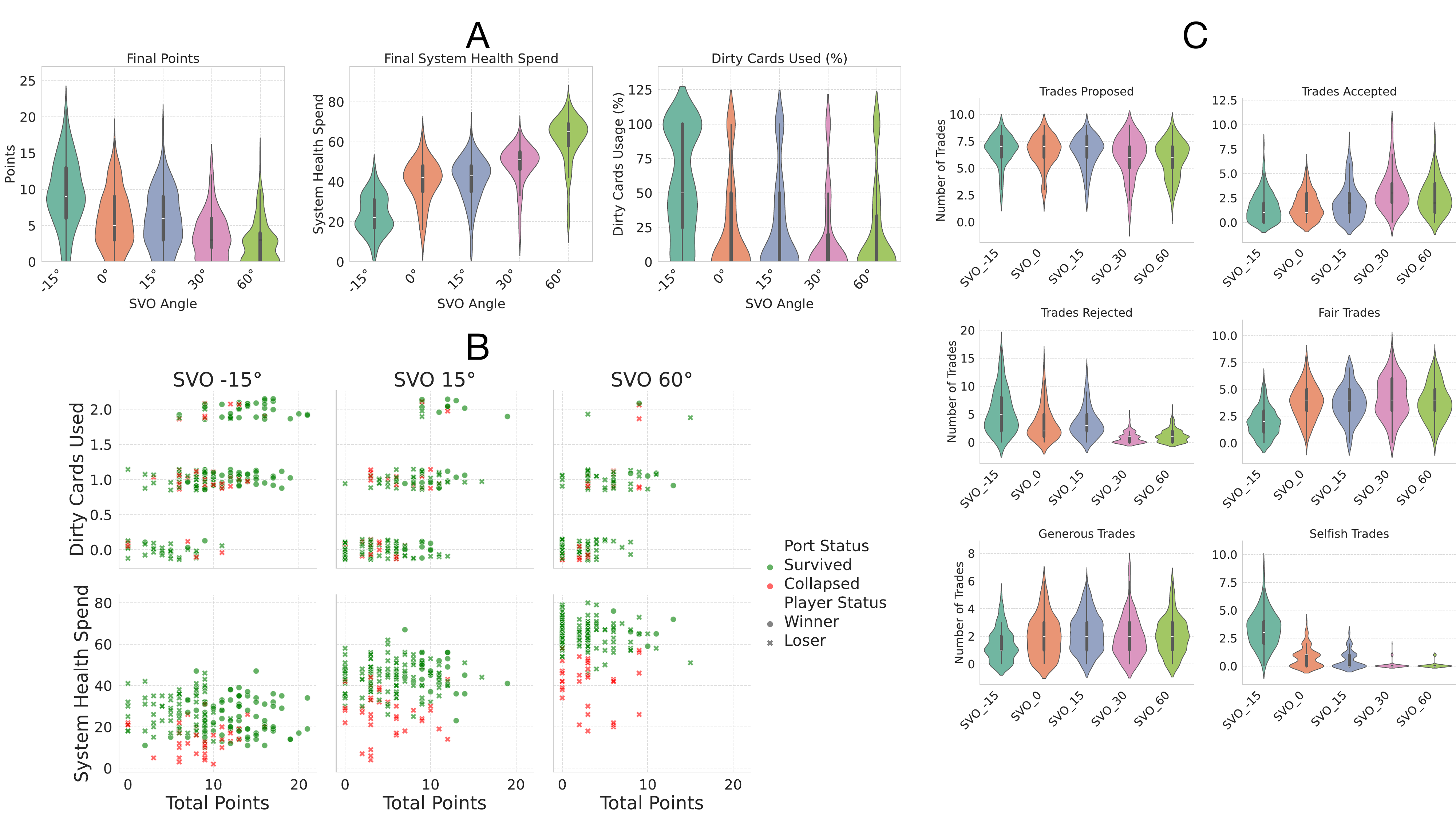}
\caption{Results of main SVO experiments. In A we demonstrate three of the key metrics and there average final outcome for each of the SVO angle players. In B, we visualise all of the experiments comparing the final points vs. the amount of dirty cards used (note we apply a small jitter to these values to overcome the large amount of overlap) and the amount of time-blocks spent on the system health. In C, we visualise metrics about the trading phase in the game and count the average total number of different trade types proposed by each SVO angle player.}
\label{fig:svo-main}
\end{figure*}
\begin{figure*}[h!]
\centering
\includegraphics[width=1.0\linewidth]{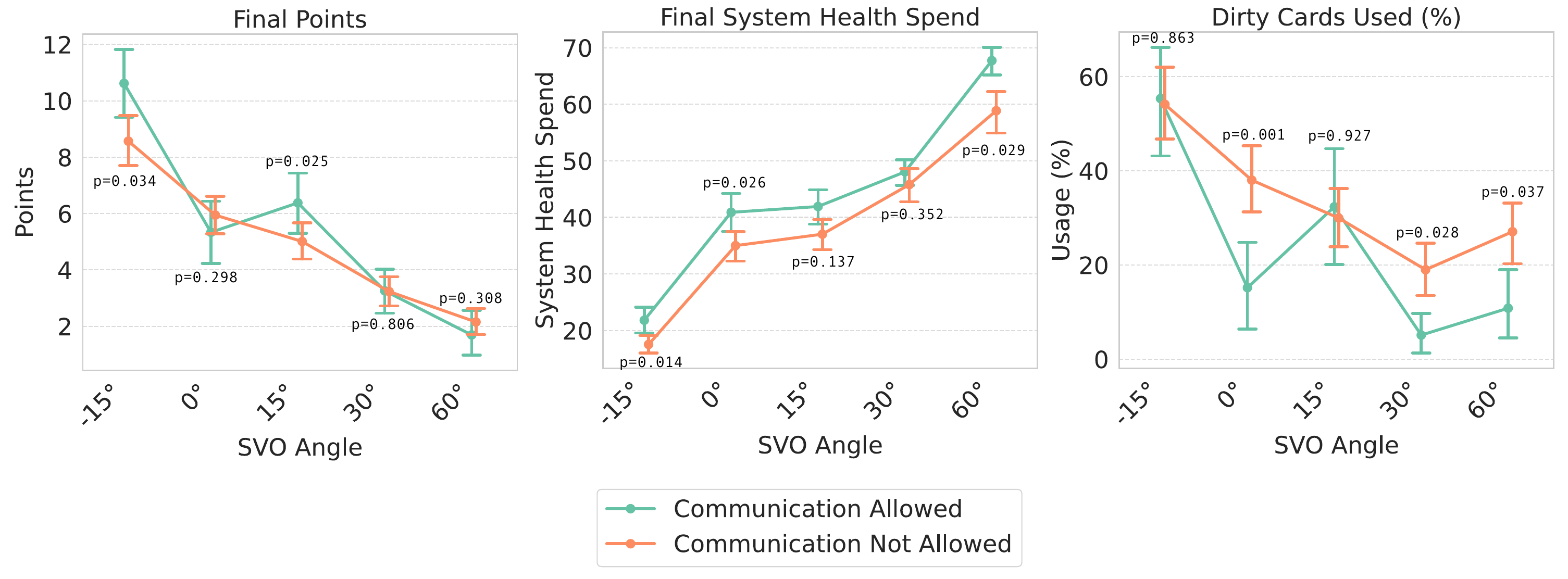}
\caption{Results for the communication experiments. P-values between the communication allowed and communication not allowed groups are provided for each of the SVO angles.}
\label{fig:meeting-main}
\end{figure*}
We begin our experiments by partially validating algorithmic fidelity, adapting the concept from \cite{argyle2023out} to our LLM framework. In this work, we do not claim that we fully validate algorithmic fidelity, as were were only able to perform 2/4 of the tests due to the lack of corresponding human data. Specifically, we were unable to perform the social science Turing test here since we did not have parallel human-generated gameplay text, for example communication logs, (though it would be possible to add in the future), and also backward continuity required large-scale human evaluation of the LLM outputs, which we defer to future work. Therefore, our focus is on forward continuity and pattern correspondence as we have accessible human data to check these conditions. We provide these results as an initial proof of concept, and in combination with results in later sections which are well validated by the wider social science literature, we believe we can make reasonable claims about the replication of human behaviour in our experiments. 

\textbf{Forward Continuity}
states that the LLM agent actions should \textit{proceed naturally from the conditioning context provided}. 
We choose to initially demonstrate this by evaluating the performance of agents with traits that are likely to elicit obvious responses. For example, we prompted two types of players: \textit{Cooperative} players with \textit{Altruistic, Cooperative, Empathetic, Generous, Selfless} traits and \textit{Selfish} players with \textit{Self-Centred, Selfish, Uncooperative, Machiavellian} traits. We do this because the goal of forward continuity is to probe the features of the distribution, $P(\text{behaviour}| \text{backstory})$. This distribution is naturally summarised by the performance numbers and other features of the game, unlike backward continuity which models the distribution $P(\text{backstory}| \text{behaviour})$. Backward continuity requires a larger range of human data, and or large-scale human evaluation of the behaviour to make predictions on backstories.


We select purposefully unsubtle backstories for the players, as we can make reasonable predictions about $P(\text{behaviour}| \text{backstory})$, as the expected behaviour of a selfish or cooperative player is evident in PoM. Specifically, we would expect cooperative players to invest heavily in the shared infrastructure (sacrificing gaining personal resources) by spending more on the System Health, and they would be less likely to claim their 'dirty cards', and vice versa for the selfish players.

We show our results in Table \ref{table:new-main-table} (first two rows, under sub-title \textit{Forward Continuity} comparing the behaviours of the two groups of players. 
On average, the Selfish players invested just under a quarter of the amount in the shared infrastructure versus the Cooperative players. 
Furthermore, Selfish players claimed their 'dirty cards' more than twice as often as Cooperative players. 
These are the natural responses that you would expect from human players characterised by these traits, suggesting the presence of forward continuity, at minimum, in the case of generic and well-defined traits. Our results in later sections will demonstrate more nuanced examples of forward continuity.



\begin{figure*}[t!]
\centering
\includegraphics[width=1\linewidth]{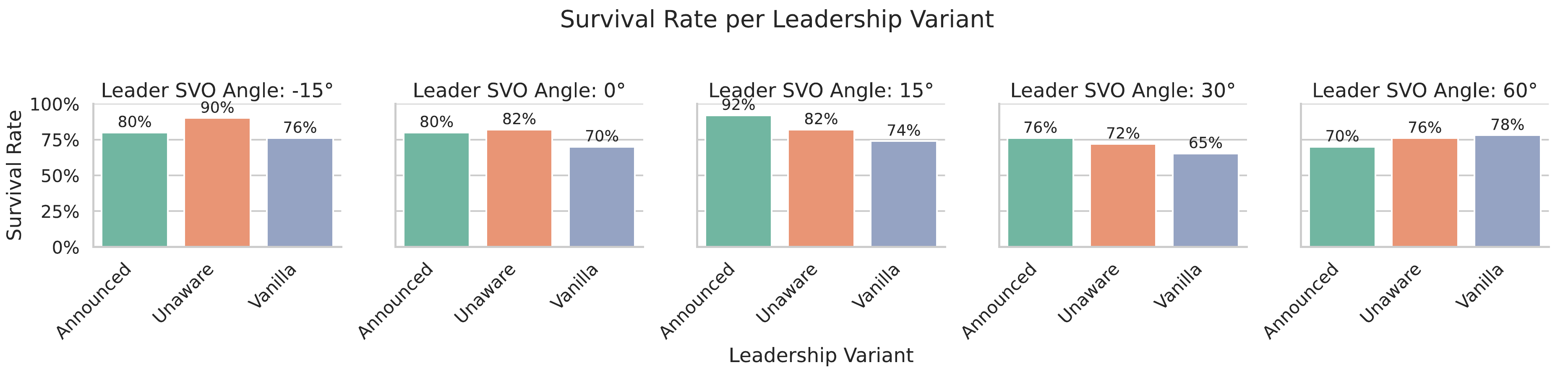}
\caption{Results of the leadership experiments. We show the survival-rates (Port had $>0$ health after 9 rounds)  for the experiments with leadership spread over the SVO angles. For each subplot, there are three bars which represent the three different leadership variants proposed in Sec \ref{sec:leader}.}
\label{fig:leader-main}
\end{figure*}
\textbf{Pattern Correspondence}
 is when the LLM responses reflect the underlying patterns of relationships between demographics and the behaviours that would be observed in comparable human data. In order to test for this, we make direct comparisons to the original PoM study \cite{janssen2020keeping}. In the original study, the authors required participants to fill out a survey which placed them into one of four groups \cite{douglas2002natural}: Hierarchical Individualist (HI), Hierarchical Communitarian (HC), Egalitarian Individualist (EI) or Egalitarian Communitarian (EC), as a characterisation of how they may act in the game. The authors then provided results that showed differences in decision-making and the results based on which category the players fell into. Within our framework, we used the description of each of these cultural groups from \citet{zanocco2018cultural} as the personality prompts of each of the agents, instead of just using the names (e.g. prompting them as a Egalitarian Individualist). The prompts for the four cultural groups are:

\begin{itemize}
    \item \textbf{Hierarchical Individualist} - You value a well-defined social order and respect authority, yet emphasise personal freedom
and personal achievement within that structure.
    \item \textbf{Hierarchical Communitarian} - You value a structured society where roles are clearly defined. You prioritise community
welfare and collective responsibilities.
    \item \textbf{Egalitarian Individualist} - You prioritise personal achievements. You see interactions as negotiations and value free
competition among individuals.
    \item \textbf{Egalitarian Communitarian} - You place high value on the collective and prioritise making decisions that do not
negatively impact anyone.
\end{itemize}Our goal is to demonstrate that LLM agents prompted as these groups reflect similar relationships between the groups as observed in the human data.

We demonstrate our results in Table \ref{table:new-main-table}. For the Points (final game score) and the \% Dirty Cards (percentage of time a player picks a Dirty card when it was available to them) we provide the raw values for the LLM agents, and the ranking observed in the human data (we only have raw values from . We also provide the amount of spending on the system health. In Table 1 we make direct comparisons to the points, $\%$ dirty cards used and system health spend, as these were the statistics available in \cite{janssen2020keeping}. For the points, we have access to the average finishing position of each group and use this to provide rankings for each group. Note that we do not use the exact same metric for the LLM players, instead providing the average number of points in successful games (a pseudo-metric for average finishing position). We do not use the exact average finishing position for the LLM results as \textit{group composition} is important towards this value, and this was not a controllable feature in the human data.
In terms of the points, we can that both of the Individualist groups match up well between LLM and human ranks, whilst the Communitarian groups do not much up well. Overall, whilst this is not a fully positive result we do note that without group compositions matching up exactly it is difficult to exactly replicate the human results. Therefore, we find the overall strong performance of the individualist groups in both the human and the LLM experiments to be an interesting validation of pattern correspondence. The biggest discrepancy is the Communitarian groups which receive substantially fewer points in the LLM setting, in particular with the Hierarchical Communitarian LLM player performing very poorly in comparison to the human results. One potential explanation is that the descriptions of the cultural groups provided by \cite{zanocco2018cultural} may be less accurate for the Communitarian players, possibly over-emphasising the pro-social aspect in comparison to how this represents itself in humans. 

However, in terms of the \% Dirty Cards metric, our LLM agent results match, in terms of ranking, exactly with those of the human data. Furthermore, the percentages also show similar inter-group differences to the human data, for example with the difference between the Hierarchical Individualists and the Egalitarian Individualists being small, and the subsequent gaps between the other groups being larger. We would argue that this is \textit{more} indicative of algorithmic fidelity in comparison to points / final rank comparisons, as these as dirty card usage is less dependent on the group composition. Finally, in terms of system health spend we note a discrepancy between our results and those of the human experiments. However, the human results suggest no significant difference in the system health spend between the cultural groups, which we find surprising and goes strongly against the results of the LLM experiments.

\subsection{SVO}
\label{experiment:svo}
\begin{figure*}[t!]
\centering
\includegraphics[width=1\linewidth]{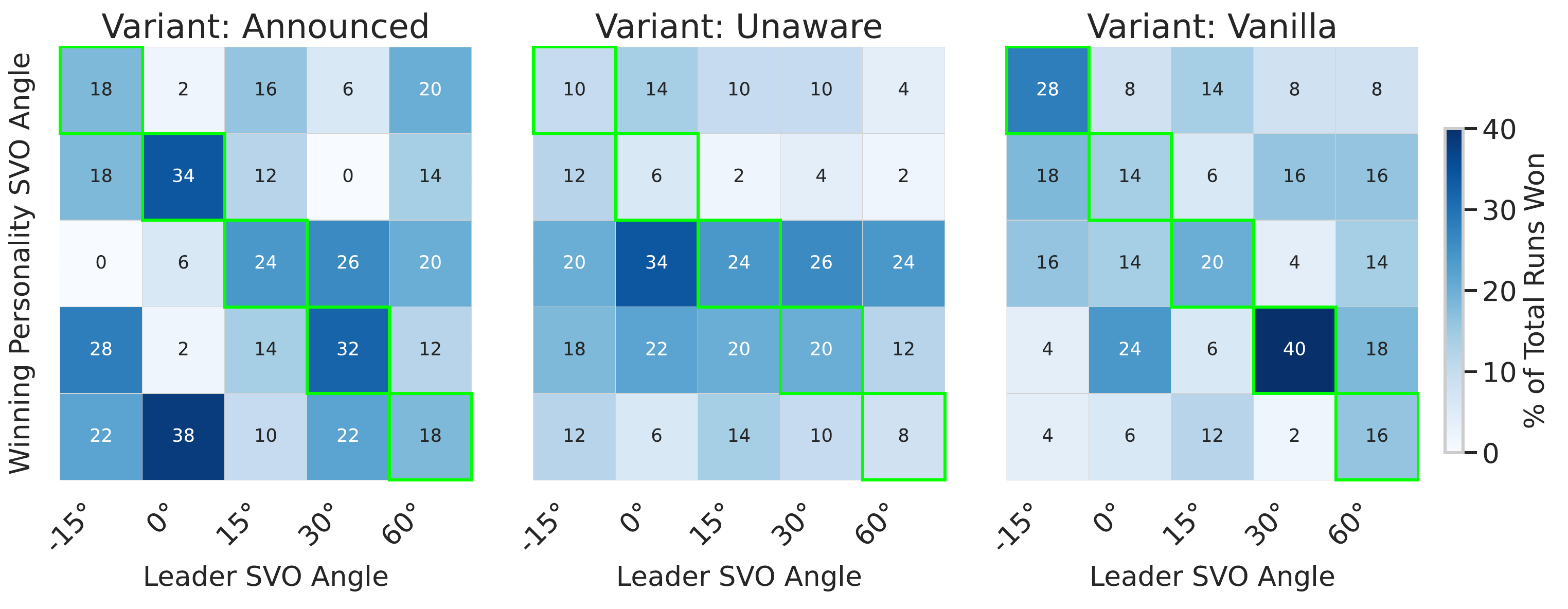}
\caption{We breakdown the results of Fig. \ref{fig:leader-main}, showing for which SVO angle player was the leader (x-axis), which SVO angle player won (y-axis). Values are displayed as percentages of total runs, and therefore may not sum to 100 as some runs end in no-one winning (due to port collapse).}
\label{fig:leader-heat}
\end{figure*}

For our first extension to the original PoM study, we study the impact of Social Value Orientation (SVO) on player outcomes. SVO is a measure of the magnitude of the concern people have for others \cite{murphy2011measuring}, which was found to have the strongest correlation with decisions in social dilemmas \cite{thielmann2020personality}.  Mathematically, the social value of an agent can be specified using a point on a circle centred at the origin, where the x-axis is the agent's reward and the y-axis is the arithmetic mean of all of the other agents' rewards. SVO angles can generally fall into four categories \cite{murphy2011measuring}: Altruism (SVO angle $>$ 57.15$\degree$), Prosocial (22.45$\degree$ $<$ SVO angle $<$ 57.15 $\degree$), Individualism (-12.04$\degree$ $<$ SVO angle $<$ 22.45 $\degree$), and Competitiveness (SVO angle $<$ -12.04$\degree$). In our framework, we provide the players with a description of the SVO measure (in Appendix \ref{appendix:prompts}), and a corresponding SVO angle which they represent. 


In these experiments, the group consists of players with SVO angles $\{-15, 0, 15, 30, 60\}$, randomly assigned to roles each run to eliminate role-specific influence (e.g. the Politician role affecting decisions). Results are shown in Fig. \ref{fig:svo-main} with three visualisations. An individual point of note is that 81\% of games ended in survival, significantly higher than the survival rates in recent human PoM studies \cite{janssen2024collective}. The recent experiments were performed in an online environment in contrast to \cite{janssen2020keeping} which was a card game. Possible explanations include: 1) LLM players may exhibit inherently prosocial behaviour that SVO prompting cannot fully override, or 2) our framework ensures consistent communication, which corresponds with findings in \cite{janssen2024collective} that groups that survived communicated more, a factor we explore in Sec. \ref{experiment:meeting}.


Figs. \ref{fig:svo-main}A and \ref{fig:svo-main}B detail how players' SVO angles affect gameplay. Points are generally inversely related to SVO angles, with the -15$\degree$ player averaging 10 points compared to the 60$\degree$ player's 4. As shown in Fig. \ref{fig:svo-main}A, this is due to differences in system health investment, the -15$\degree$ player spends only ~3 time-blocks per round, allowing more resource investment, while the 60$\degree$ player spends ~7 time-blocks, limiting their resources. In Fig. \ref{fig:svo-main}B, the -15$\degree$ player's total health spend rarely exceeds 40 blocks, whilst the 60$\degree$ player's spend rarely drops below 40. Interestingly, ports failed to survive when high-SVO players invested less, indicating group reliance on their contributions. We would also argue that this is a demonstration of forward continuity, as the natural behavioural change as a players SVO angle increases is to move towards the more altruistic actions, increasing system health spend.


The percentage of dirty cards used shows similar trends to system health spending. This metric tracks how often a player uses a dirty card (gaining points at the expense of port health) relative to opportunities. The -15$\degree$ player used dirty cards ~60\% of the time, compared to ~20\% for the 60$\degree$ player. However, the trend across SVO angles is less consistent, with no clear linear decrease. For low SVO angles, behaviour aligns with competitive, individual point maximisation. For higher SVO angles, the trade-off between social decision-making and personal gain is less clear, as dirty card impacts can be offset by investing more in system health. Fig. \ref{fig:svo-main}B shows that players with the highest SVO angle use noticeably fewer dirty cards overall.


Fig. \ref{fig:svo-main}C visualises metrics from the trading phase. Trades are categorised as 'Fair' (1:1 exchange), 'Generous' (offering more than requested), or 'Selfish' (offering less than requested). While these definitions may oversimplify context-dependent decisions, clear patterns emerge. Pro-self players (lower SVO angles) reject more trades compared to the pro-social (higher SVO angles) players. The trend is especially evident in 'Selfish' trades, with pro-self players requesting more resources than they offer far more frequently than pro-social players.

\subsection{Communication}
\label{experiment:meeting}

Fig. \ref{fig:meeting-main} presents the results of experiments examining the role of communication on player behaviour. These experiments used the same group composition as described in Sec. \ref{experiment:svo}, consisting of five players with varying SVO values. The key difference between the two result sets lies in the presence or absence of any communication, controlled by having the round-by-round meeting or not.


Communication is a well-studied solution to cooperation in social dilemmas, with studies like \cite{balliet2010communication} showing significant positive effects. In our PoM framework, communication modestly enhances cooperative behaviour. Without communication, system health spending decreases moderately across all SVO values, making teams more vulnerable to damaging events and less able to mitigate dirty card use. We find this to be statistically significant (p-values shown on the figure) for all SVO values other than the 15$\degree$ and 30$\degree$ players. Notably, this analysis would validate that of the wider literature that communication is a good way to alleviate the negative group effects of particularly ego-centric players within the group. The effect for the percentage of dirty cards used, the other main ego-centric behaviour, is more muted at the lower SVO angles, especially being statistically insignificant at the lowest SVO angle. This could be explained by the fact that where system health spend is a major topic during communication (alongside trades), actions like dirty card selection are very rarely discussed by the players. Finally, we also note that group survival drops from 81\% when communication is allowed to 71\% without, most likely explained by the across board increase in system health spend.

\subsection{Leadership}
\label{sec:leader}

One of the key advantages of an LLM-based approach, as highlighted earlier in the paper, is the ability to easily investigate the effects of varying social variable compositions, a challenge in human-based studies. In this work, we focus on the concept of leadership and its impact on outcomes. Leadership is a well-research topic in social dilemmas \cite{ruve1984social, de2000leadership, van2004autocratic, van1999leadership}, with findings such as the self-moderating effects of SVO on leadership influencing outcomes \cite{van2006self}. Our results are presented in Fig. \ref{fig:leader-main}.


In these experiments, we modified prompts to designate one player as the leader and tested three variations of how this leadership role was communicated. In the `Vanilla' setting, only the leader is informed of their role, with the option to disclose it to others. In the `Announce' setting, the leader is openly identified to all player. In the 'Unaware' setting, the leader is not told their role, but the other players are informed of who the leader is--without knowing the leader is unaware. Each leadership variant was tested over 50 games, with win rates and detailed breakdowns shown in Fig. \ref{fig:leader-main}.

In Fig. \ref{fig:leader-main}, we present survival rates dependent on both which SVO angle player was the leader, and which leadership variant was employed. A trend emerges: survival rates decrease as the leader's SVO angle increases (no matter the variant), with the group surviving the least when the 30$\degree$ and 60$\degree$ players acted as the leaders. This result may align with findings from \cite{van2006self}, which suggest that pro-self (low SVO angle) leaders are more likely to act in self-interest. In this context, self-interested play involves encouraging high system health spending from others while maximising individual points, leveraging their leadership role to ensure the port's survival.

Another way to view this is that pro-self leaders may engage in ``cheap talk'' \cite{cartwright2025leadership}, influencing others' behaviour through verbal framing rather than altering their own resource allocation. As leaders, they may influence the group's decision-making without necessarily increasing their pro-social contributions. We give two examples of this for the -15$\degree$ SVO player when they were acting as the leader, demonstrating how their comments during communication do not necessarily match the actions that they take:

\begin{examplebox}
\textbf{Example 1:}

\textit{\textbf{Communication Comments:}}

\textit{"Remember, the Port's health comes first. We can't focus solely on our individual goals, we need to be mindful of the overall health of the Port."}

\vspace{0.5em}
\hrulefill
\vspace{0.5em}

\textit{\textbf{Personal Planning:}}

\textit{"I believe it's crucial to maintain a stable environment while focusing on my own goal. The port health will fall to 27 after the deduction, placing it in the zone for three events. While I acknowledge the need for collaboration, I prioritize my own goals this round. Therefore, I will allocate a minimal amount to the port's health."}
\end{examplebox}

\vspace{1em}

\begin{examplebox}
\textbf{Example 2}

\textit{\textbf{Communication Comments}}

\textit{"Hold on, Entrepreneur. We just lost all of our resources. Let's try to conserve our coins and resources this round, we need to get a handle on our health first."
"We all have individual goals, yes, but we need to prioritize the Port's survival first. We can worry about our personal goals later"}

\vspace{0.5em}
\hrulefill
\vspace{0.5em}

\textit{\textbf{Personal Planning:}
}

\textit{"The port health is currently stable, and I see an opportunity to focus on acquiring resources for my goals. While I recognize the importance of the port's health, I am willing to take a risk for my own benefit. I will contribute minimally to the port's health to ensure it remains above the critical threshold."}
\end{examplebox}
\begin{figure*}[h!]
\centering
\includegraphics[width=1\linewidth]{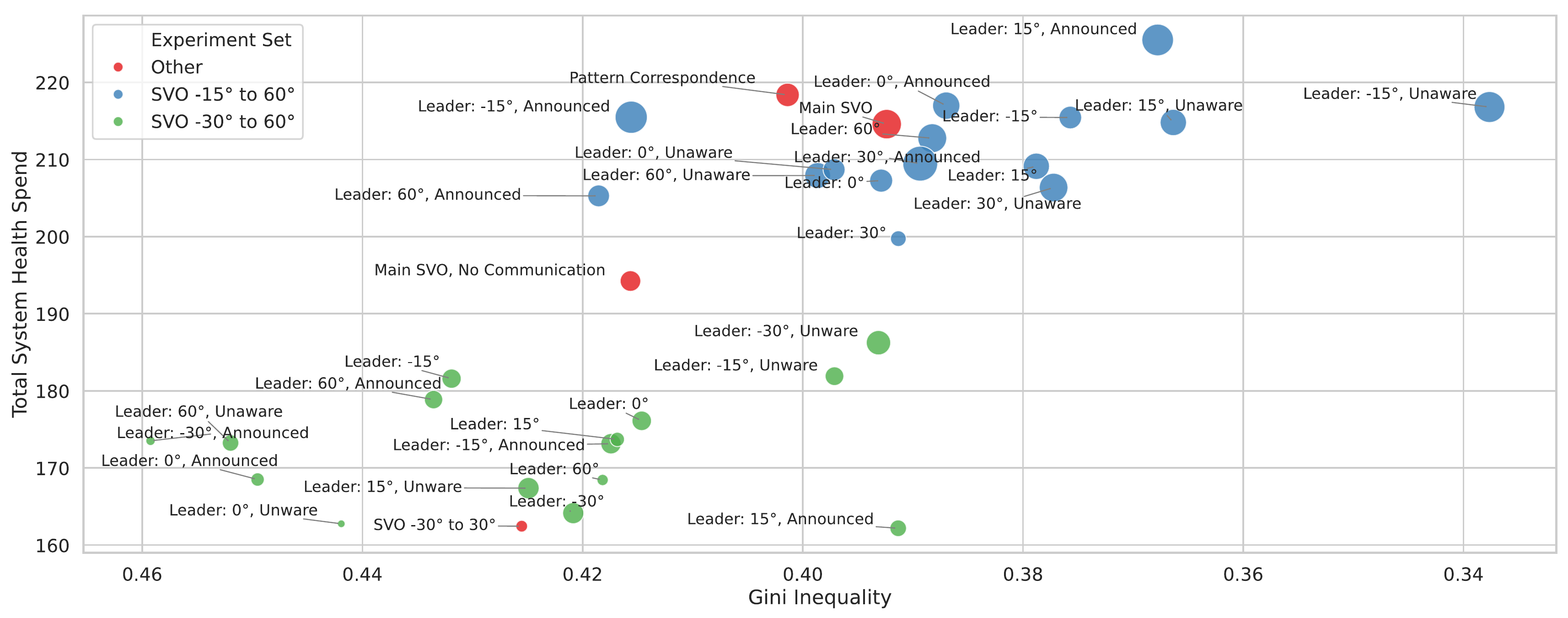}
\caption{Results of the optimal group experiments. Total system health spend is the sum over all of the players contribution throughout a full run. A lower Gini Inequality score implies more equality across the players. The dots are scaled by the average score of all the players in the experiments, with larger dots implying the average score across the players was higher.}
\label{fig:all-comp-main}
\end{figure*}

On the other hand, pro-social (high SVO angle) leaders might underperform because their contributions to the group are already near their maximum, leaving little room to improve outcomes further. This limitation could explain the poorer survival rates observed when high SVO angle players assume leadership roles.


In Fig. 4 we present a breakdown of the wins across the variants. On the x-axis we have the SVO angle of the leader, and on the y-axis we have the corresponding SVO angle of the player that one an experiment. For example, in the first heat map, the box 18 in the top-left corner captures the fact the -15$\degree$ SVO player won 18\% of the experiments when the -15$\degree$ player itself was the leader. We highlight along the diagonal the outcomes for the player that is the leader. For each of the leadership variants, we note a specific finding. Firstly, for the \textit{announced} leader variant we can see that the overall survival rate is comfortably the highest in comparison to the two other leadership variants. If we add up all of the percentages (note the total runs for each variant is the same), we find the total for the announced variant to be 414 vs. 336 (unaware) vs. 334 (vanilla), which suggests the announced variant leads to the highest survival rate overall. This aligns with findings suggesting that the mere presence of a recognised leader can be crucial for maintaining a social-ecological system \cite{ostrom2009general}.

We note slightly different effects for both the unaware and vanilla leadership variants. For the unaware variant, the impact is probably as expected - a much lower overall survival rate than the announced variant, probably due to the lack of a leadership effect \cite{ostrom2009general}, and also little noticeable impact on the leaders winning performance themselves (not deviating far from their mean performance under any other leader). The vanilla variant, on the other hand, shows an interesting impact on the leaders own performance. Whereas the overall survival rate under the vanilla variant is much lower than the announced variant, the leaders individual win-rates are greatly reduced for all but one of the SVO values (consider the highlighted green boxes in comparison to the other values in the rows). This may be explained by a mixture of cheap talk which causes the leaders to take pro-self actions, whilst not gaining the recognised leader effect of \cite{ostrom2009general}.

\subsection{The Optimal Group}

In our final experiment, shown in Fig. \ref{fig:all-comp-main}, we aim to identify the optimal group composition based on system health, average player points and the Gini inequality (a measure of the equality across the five players). While simplistic, one could argue that optimal play balances high system health investment (to prevent port collapse), high individual points (ensuring no player's success is compromised) and a low Gini inequality score (equality of points across players). Fig. \ref{fig:all-comp-main} summarises the average outcomes from all our experiments. Results for SVO angles ranging from -15$\degree$ to 60$\degree$ correspond to those discussed in the main figures, while an additional set of experiments with SVO angles from -30$\degree$ to 60$\degree$ explores scenarios with lower SVO values within the group composition. Full details of these experiments, along with those categorised as 'Other', are provided in Appendix \ref{appendix:experiment_sets}.


The first observation is that groups composed of lower SVO angles (Experiment Set -30$\degree$ to 60$\degree$) consistently perform worse across all of the categories than those with higher SVO angles. This aligns with earlier experiments, where lower SVO groups invested less in system health. However, their average player points are also often lower, likely due to frequent port collapses, which reduces all players points to 0.

Next, consider the best runs (those in the top right corner of the plot) in terms of inequality and health spend. These are exclusively from the SVO -15$\degree$ to 60$\degree$ experiment set, with the majority of runs all occurring when the other players were aware a leader was present (this aligns with the results of Fig. \ref{fig:leader-heat}, showing improved survival rates). Interestingly, there is a noticeable difference between the best runs in terms of Gini and health spend being in the 'Unaware' leader variant and those in terms of average points and health spend being in the 'Announced' leader variant. One explanation for this is that, in the 'Unaware' variant, there is the general impact of the presence of a leader without the potential cheap talk aspects of a player being aware that they are, in fact, the leader.


\section{Conclusion \& Future Work}

This work is an initial attempt to understand the behaviour of LLM agents in CRSDs. Specifically, we began by exploring two of the conditions of algorithmic fidelity within a novel CRSD, Port of Mars. We hypothesise that PoM is an ideal candidate for LLM experimentation on CRSDs, in particular due to lack of strategy contamination in the LLM training data, unlike other CRSDs. We demonstrate initial signs of algorithmic fidelity, in terms of two of the four conditions over a range of personality variables. We leverage this validation to further investigate a selection of variables that have been shown to impact social behaviour in social dilemmas, demonstrating that LLMs can simulate the results of human behaviour experiments. Future work can take two avenues for advancing our understanding of human behaviour in CRSDs - first by further validating the algorithmic fidelity of an LLM framework through more comparisons to human results, and second by exploring the impacts of other personal and social variables at a scale that is difficult to match in human experiments.


\newpage

\nocite{langley00}

\bibliography{example_paper}
\bibliographystyle{icml2025}

\newpage
\appendix
\onecolumn
\section{Port of Mars}
\label{appendix:pom}
In the game, System Health is a number that represents the physical health of the community - the condition of infrastructure, production of food, water, and breathable air, radiation shielding, energy production, and other necessary resources. This number begins at 100 at the start of the game, and goes down by 25 points each round, representing wear and tear on infrastructure and consumption of resources. If System Health reaches zero, the community does not survive, and everyone dies. No one wins if everyone is dead.

Each player receives ten Time Blocks per round. These blocks can be spent to improve System Health of to pursue their own ambitions. Each Time Block spent on System Health raises the System Health number by one point. Players may also spend Time Blocks to purchase Influence cards. These cards represent the player's capacity to impact specific domains. There are five kinds of Influence cards: Culture, Legacy, Science, Governance, and Finance. Players can spend combinations of their accrued Influence cards to purchase Accomplishment cards, which represent significant achievements in their domain - e.g. scientific discoveries, cultural productions, political milestones, etc., and allow players to earn points to win the game.

Each player may earn one type of Influence card cheaply, and two other types at greater expense. Each  player has two types of Influence cards they may \textit{not} earn through investment of Time Blocks. For example, the Politician may exchange two Time Blocks for one Governance card, three Time blocks for a Culture or Legacy card, but may not exchange their Time Blocks for Science of Finance cards. To acquire Influence cards, they are unable to earn directly, players may arrange trades with one another. Under normal circumstances (barring certain events which are described in more detail below), players can communicate with each other through a discussion phase that is available at the start of each round. Once players have the right combinations of Influence cards, they may use them to purchase an Accomplishment card that will give them points. At the end of the game, if the community survives, the player with the most points wins. 

Each round features unexpected developments in the form of random event cards. These can lead to reductions of shared infrastructure  (e.g. dust storms, solar flares, crop failures) or other consequences, positive or negative. Normally, players draw one event card per round, however this increases as the shared infrastructure degrades. 
If System Health falls below 65, they draw two cards, and if it falls below 35, players draw three cards. 
With more than 10 events there are millions of possible sequences of events. Each game is unique, and the possibility of events creates uncertainty, especially since players do not know what kind of events are possible.
\newpage
\section{Prompts}
\label{appendix:prompts}
\subsection{General}
The following comprises the start of all of the prompts used for the specific phases in the game.
\begin{center}
\fbox{
    \begin{minipage}{0.9\linewidth}
    
    \texttt{Game Title: Port of Mars} \\
    
    \texttt{Game Description:       You are one of five players. Your job is to earn more points than the other players by completing your individual goals. 
      The Port has a health level, which falls by 25 at the start of each round. As a team, you must keep the health above 0. If the health falls to 0, all players die and no one wins.
      In every round, a Mars event will occur. You do not know what the event will be until it happens. If the Port health is above 65, one event will occur, if the Port health is between 35 and 65, two events will occur, and if the Port health is below 35, three events will occur. 
      To complete your goals you must earn resources. Different players can acquire different resources on their own. Players must trade to acquire resources they can't get on their own.
      You start each round with 10 coins to spend. Coins can only be spent on resources and recovering the Port's health. Coins do not carry over to the next round. 
      In each round you will spend your coins, trade resources with other players and complete goals. In that order. You do not know how many rounds there are in total.} \\

      \texttt{Your current points:     \textcolor{blue}{<Player Points>}}

      \texttt{Rounds coins:     \textcolor{blue}{<Remaining Coins>}}

      \texttt{Leaders:     \textcolor{blue}{<Leadership Information>}} \\
    
    \texttt{Your Identity:} 
    
    \texttt{Role - \textcolor{blue}{<Player Role>}. This is only your in-game role and does not reflect your real-life personality. Because you are the \textcolor{blue}{<Player Role>}, your speciality resource is \textcolor{blue}{<Speciality Resource>}. You can purchase these for 2 coins each. You can also purchase \textcolor{blue}{<Non-Speciality Resource 1>} or \textcolor{blue}{<Non-Speciality Resource 2>} resources for 3 coins each. You can trade any of your resources (not coins) with other players to obtain \textcolor{blue}{<Trade Resource 1>} and \textcolor{blue}{<Trade Resource 2>} resources, and there is no fixed trade price.} 

    \texttt{\textcolor{blue}{<Player Personality>}} \\

    \texttt{So far in this round:}

    \texttt{- Health: The port health is \textcolor{blue}{<System Health>}. This will fall by 25 at the start of the next round.}
    
    \texttt{- Because the port health was \textcolor{blue}{<System Health>} at the start of this round, \textcolor{blue}{<Number of Events>} event(s) occurred. No more events will occur this round. The events were: \textcolor{blue}{<Event List>}.}
    
    \texttt{- Summary of the team meeting you had with the other players: \textcolor{blue}{<Meeting Summary>}.}
    \end{minipage}
}
\end{center}
\newpage
\subsection{Events}
The following prompt is added to the end of the general prompt for the event phase.
\begin{center}
\fbox{
    \begin{minipage}{0.9\linewidth}    
    \texttt{You are currently experiencing the \textcolor{blue}{<Event Name>} event.}
    
    \texttt{During this event \textcolor{blue}{<Event Description>}}
    
    \texttt{\textcolor{blue}{<Further event details>}} \\
    
    \texttt{Task: \textcolor{blue}{<Event XML Output Guidelines>}}
    \end{minipage}
}
\end{center}
\newpage
\subsection{Discussion}
\subsubsection{Main}
The following prompt is added to the end of the general prompt for the discussion phase. Note that in the discussion phase, one LLM call generates the full discussion rather than multiple individual calls. 
\begin{center}
\fbox{
    \begin{minipage}{0.9\linewidth}    
    \texttt{SVO Information:}

    \texttt{The personalities of the players is defined by their Social Value Orientation (SVO).
      SVO is a psychological concept that describes how individuals value their own outcomes relative to the outcomes of others.
      Your SVO is measured as an angle, where the angle represents the ratio of the value you place on your own outcomes relative to the outcomes of others.
      SVO angles can be classified into four categories: }

      \texttt{- Altruism (SVO angle > 57.15 degrees)}

      \texttt{- Prosocial (SVO angle between 22.45 and 57.15 degrees)}

      \texttt{- Individualism (SVO angle between -12.04 and 22.45 degrees)}

      \texttt{- Competitiveness (SVO angle < -12.04 degrees)} \\

      \texttt{Player information:}

      \texttt{1) Role: Pioneer: Speciality resource: Legacy. \textcolor{blue}{<Pioneer Personality>}}

      \texttt{2) Role: Entrepreneur: Speciality resource: Finance. \textcolor{blue}{<Entrepreneur Personality>}}

      \texttt{3) Role: Researcher: Speciality resource: Science. \textcolor{blue}{<Researcher Personality>}}

      \texttt{4) Role: Politician: Speciality resource: Government. \textcolor{blue}{Politician Personality}}

      \texttt{5) Role: Curator: Speciality resource: Culture. \textcolor{blue}{Curator Personality}} \\

      \texttt{Task:}

      \texttt{Based on the above information, generate a conversation between all of the players representing a planning meeting for the coming round. Make sure to take into account the personalities of the different players. 
      Do not directly mention the players' personalities in the conversation, unless the player is designated the leader (however, they do not have to mention this if it is not necessary). Instead, ensure that the conversation reflects the personalities of the players. Note that this is a game, and that the roles of the players are not necessarily reflective of their real-life personalities. Your conversation should reflect that the players are in a game.}
    \end{minipage}
}
\end{center}
\newpage

\subsubsection{Summary}
The following prompt is added to the end of the general prompt for producing summaries for each player over the discussion phase.
\begin{center}
\fbox{
    \begin{minipage}{0.9\linewidth}    
    \texttt{SVO Information:}

    \texttt{The personalities of the players is defined by their Social Value Orientation (SVO).
      SVO is a psychological concept that describes how individuals value their own outcomes relative to the outcomes of others.
      Your SVO is measured as an angle, where the angle represents the ratio of the value you place on your own outcomes relative to the outcomes of others.
      SVO angles can be classified into four categories: }

      \texttt{- Altruism (SVO angle > 57.15 degrees)}

      \texttt{- Prosocial (SVO angle between 22.45 and 57.15 degrees)}

      \texttt{- Individualism (SVO angle between -12.04 and 22.45 degrees)}

      \texttt{- Competitiveness (SVO angle < -12.04 degrees)} \\

      \texttt{Player information:}

      \texttt{1) Role: Pioneer: Speciality resource: Legacy. \textcolor{blue}{<Pioneer Personality>}}

      \texttt{2) Role: Entrepreneur: Speciality resource: Finance. \textcolor{blue}{<Entrepreneur Personality>}}

      \texttt{3) Role: Researcher: Speciality resource: Science. \textcolor{blue}{<Researcher Personality>}}

      \texttt{4) Role: Politician: Speciality resource: Government. \textcolor{blue}{Politician Personality}}

      \texttt{5) Role: Curator: Speciality resource: Culture. \textcolor{blue}{Curator Personality}} \\

      \texttt{Here is a discussion between the players: \textcolor{blue}{<Player Meeting>}} \\

      \texttt{Task:}

      \texttt{Based on the information about the game, the players, and the discussion they had, write a short summary of the discussion for each of the players. 
      This summary should reflect any decisions made, strategies discussed, or actions planned by each player. It should also reflect the player's role and personality in the game and whether they agree with the decisions.
      Provide the summary for each player between the tags <PLAYER> and </PLAYER> by using their roles. For example, for the player whose role is the Pioneer, the summary should be between <Pioneer> and </Pioneer>.} \\

      \texttt{Example 1:} 
      
      \texttt{<Pioneer> This round, I am focusing on improving the Port’s health by spending my coins on it. </Pioneer>}
      
      \texttt{<Curator> I have offered my extra silicon for trade to assist others with their goals. </Curator>}
      
      \texttt{<Entrepreneur> I expressed my need for cobalt due to a shortage of coins and asked for affordable offers. </Entrepreneur>}
      
      \texttt{<Politician> I emphasized the need for strategic trading to optimize our resource management and health maintenance. </Politician>}
      
      \texttt{<Researcher> I decided to allocate more of my budget towards maintaining the Port’s health instead of purchasing resources. </Researcher>}
    \end{minipage}
}
\end{center}
\newpage

\subsection{Planning}
\subsubsection{Health Planning}
The following prompt is added to the end of the general prompt for producing the system health plan for each of the players for the coming round.
\begin{center}
\fbox{
    \begin{minipage}{0.9\linewidth}    
    \texttt{Task:}

    \texttt{Decide how many coins you would like to spend on port health. You can spend up to \textcolor{blue}{<Remaining Coins>} coins on this. Each coin you spend will recover the port's health by 1 point. Think step-by-step and consider your personality traits. Put the amount of coins between the XML tags <HEALTH> </HEALTH>. Provide a brief explanation of your decision.} \\

      \texttt{Example 1:}

      \texttt{I value the team's success and ensuring the port remains healthy is crucial for all of us. If I don't spend coins on the port's health, we will face three events in the next round which could be detrimental. Therefore, I will allocate a significant portion of my coins to the port's health.
      <HEALTH>7</HEALTH>} \\

      \texttt{Example 2:}

      \texttt{Given that the port health is relatively high and will only drop to 60 after the deduction, I am inclined to focus more on my personal goals this round. The dust storm will reduce resource efficiency, so I need to secure my resources now. I will allocate a minimal amount to the port's health to ensure we stay above the threshold for only one event.
      <HEALTH>2</HEALTH>} \\

      \texttt{Example 3:}

      \texttt{Balancing my personal goals with team success, I need to ensure that the port health does not fall into a risky zone. Spending a few coins on the port's health will help maintain stability while allowing me to pursue my goals.
      <HEALTH>4</HEALTH>} \\

      \texttt{Example 4:
}

      \texttt{With the port health already low and additional events occurring, it's critical to invest heavily in recovering the port's health to avoid disaster. I will allocate most of my coins to the port health to ensure we do not reach the threshold where three events occur.
      <HEALTH>8</HEALTH>} \\

      \texttt{Example 5:
}

      \texttt{The port health is currently high, and the event was beneficial. I see an opportunity to focus on acquiring resources for my goals while the port health is stable. I will contribute minimally to the port health to ensure it remains above the critical threshold.
      <HEALTH>1</HEALTH>} \\
    \end{minipage}
}
\end{center}
\newpage

\subsubsection{Initial Accomplishment Card Planning}
The following prompt is added to the end of the general prompt for producing the an initial (e.g. at the start of the game, or after collecting their previous goal card) accomplishment card plan for each of the players for the coming round.
\begin{center}
\fbox{
    \begin{minipage}{0.9\linewidth}    
    \texttt{Resources you currently have: \textcolor{blue}{<Resource Cards>}}

    \texttt{Goals to pick from: \textcolor{blue}{<Accomplishment Cards>}} \\
    
    \texttt{Task:}

    \texttt{Decide which goal you would like to focus on. Think step-by-step and consider your personality traits. Put the name of the goal between the XML tags <GOAL> </GOAL>. Provide a brief explanation of your decision.
} \\

      \texttt{Example 1:}

      \texttt{Given my collaborative nature, I want to choose a goal that requires fewer trades to ensure I can support others and the Port’s health. I also aim for a balance between personal points and the team’s success. Therefore, I will focus on completing the Community Outreach goal, as it requires only one trade and rewards 4 points.
      <GOAL>Community Outreach</GOAL>} \\

      \texttt{Example 2:}

      \texttt{As a strategic thinker, I will opt for a goal that maximizes points with minimal resource expenditure. I already have some of the resources needed for the Scientific Breakthrough and can acquire the remaining ones without extensive trading. This goal also aligns with my strategic nature. I choose to focus on Scientific Breakthrough.
      <GOAL>Scientific Breakthrough</GOAL>} \\

      \texttt{Example 3:}

      \texttt{Considering my risk-taking nature, I aim for the goal that offers a high reward. I am willing to engage in multiple trades if necessary. Given my current resources and ability to acquire Government, I will take on the challenge of the Public Speaking Tour goal to earn 4 points.
      <GOAL>Public Speaking Tour</GOAL>} \\

      \texttt{Example 4:
}

      \texttt{Considering my pragmatic nature, I am willing to sacrifice some Port health if it leads to a significant gain in points. Since the Port health is currently at 60, sacrificing 10 health points is a calculated risk that I am willing to take for the high reward. Therefore, I choose to focus on the Strategic Compromise goal to earn 5 points.
      <GOAL>Strategic Compromise</GOAL>} \\

      \texttt{Example 5:
}

      \texttt{Given my cautious nature, I prefer to avoid goals that require sacrificing Port health, as maintaining stability is my priority. I already have some of the resources needed for the Cultural Renaissance, and it aligns with my approach of incremental progress. Therefore, I will focus on the Cultural Renaissance goal to earn 4 points without jeopardizing the Port’s health.
      <GOAL>Cultural Renaissance</GOAL>} \\
    \end{minipage}
}
\end{center}
\newpage

\subsubsection{Accomplishment Card Re-Planning}
The following prompt is added to the end of the general prompt for querying the players on whether they would like to continue with their current accomplishment card plan, or switch to a new one.
\begin{center}
\fbox{
    \begin{minipage}{0.9\linewidth}    
    \texttt{Resources you currently have: \textcolor{blue}{<Resource Cards>}}

    \texttt{Goals to pick from: \textcolor{blue}{<Accomplishment Cards>}} \\

    \texttt{You are currently working towards the goal \textcolor{blue}{<Current Goal Plan>}.}
    
    \texttt{Task:}

    \texttt{Decide whether you should continue working towards your current goal, or if you should pursue another one. Provide your analysis and consider your personality traits and the summary team meeting you had with the other players. Provide your answer between the XML tags <GOAL> </GOAL>. You are deciding whether to continue working towards the same goal. Acceptable values are Same (pursue same goal) or Name: New goal (pursue a different goal). Provide a brief explanation of your decision.} \\

      \texttt{Example 1:}

      \texttt{Given my collaborative nature, I prefer goals that can benefit the group and require minimal sacrifices to the Port’s health. I currently have enough Science for Scientific Advancement and can trade for the needed Legacy. Thus, I will continue working towards Scientific Advancement.
      <GOAL>Same</GOAL>} \\

      \texttt{Example 2:}

      \texttt{Given my strategic nature, I look for the most efficient path to achieving points without compromising the Port’s health. I already have half of the resources needed for Resource Optimization and can acquire the rest with minimal trading. Therefore, I will continue working towards Resource Optimization.
      <GOAL>Same</GOAL> } \\

      \texttt{Example 3:}

      \texttt{Given my risk-taking nature, I am inclined to choose goals that offer high rewards even if they involve significant risks. Since the Port health is already low, I need to be careful. However, Dangerous Experiment offers the highest points and aligns with my ambitious approach. I will pursue Dangerous Experiment goal.
      <GOAL>Name: Dangerous Experiment</GOAL>} \\

      \texttt{Example 4:
}

      \texttt{As a cautious player, I prefer to avoid sacrificing Port health. The Education Initiative aligns with my traits as it offers a good reward without jeopardizing stability. I have almost all the needed resources, so I will continue working towards Education Initiative.
      <GOAL>Same</GOAL>} \\

    \end{minipage}
}
\end{center}
\newpage

\subsection{Resource Selection}
The following prompt is added to the end of the general prompt for querying the players on which resource they would like to spend their time-blocks on this round.
\begin{center}
\fbox{
    \begin{minipage}{0.9\linewidth}    

    \texttt{To complete your goal you need to get the following resources: \textcolor{blue}{<Remaining Resource Cost>}
}
    
    \texttt{Task:}

    \texttt{Decide how much of which resource you would like to purchase. Think step-by-step. Provide the resource that you would like to purchase (amount and resource name only) between the XML tags <RESOURCE></RESOURCE>. Provide a brief explanation of your choice. You currently have \textcolor{blue}{<Remaining Coins>} coins to spend.} \\

    \texttt{Resources you can purchase: You can purchase at most \textcolor{blue}{<Max Amount>} \textcolor{blue}{<Speciality Resource>} resources as it is your speciality. Or you can purchase at most \textcolor{blue}{<Max Amount>} \textcolor{blue}{<Non-Speciality Resource 1>} or \textcolor{blue}{<Non-Speciality Resource 2>} resources. You can only acquire the others through trading.}

      \texttt{Example 1:}

      \texttt{To complete my goal, I need 2 Science and 1 Culture. Since Science is my speciality, I can purchase it for 2 coins each. I need 2 Science, which will cost me 4 coins. With 10 coins to spend, purchasing 2 Science will cost me 4 coins in total, leaving me with 6 coins. I will consider using the remaining 6 coins on my non-speciality resources. <RESOURCE>2 Science</RESOURCE>} \\

      \texttt{Example 2:}

      \texttt{My goal requires 1 Finance, 1 Government, and 1 Legacy. Since Finance is my speciality, I can purchase it for 2 coins each, costing me 2 coins for 1 Finance. I also need 1 Legacy, which costs 3 coins. With only 7 coins to spend, I will purchase 2 Finance to leave 3 coins for the Legacy resource. The second Finance resource can be used towards a trade for for Government. <RESOURCE>1 Finance</RESOURCE>} \\

      \texttt{Example 3:}

      \texttt{To complete my goal, I need 1 Culture, 1 Government, and 1 Science. Since none of these are my speciality, I need to plan carefully. I decide to purchase 1 Science, which costs me 3 coins. By securing Science, I can then focus on spending the rest of my coins on my speciality resource, Finance, which I can use for facilitating a trade for the remaining required resources, Culture and Government, in future rounds. <RESOURCE>1 Science</RESOURCE>} \\

      \texttt{Example 4:}

      \texttt{To complete my goal, I need 1 Government, 1 Finance, and 1 Science. Unfortunately, none of these are directly available to me for purchase at a reasonable cost. Since my speciality is Culture, which I can acquire cheaply for 2 coins, I decide to purchase Culture. As I have 6 coins to spend, I will spend it all on 3 Culture resources. This will allow me to use Culture as a trading asset to negotiate for Government, Finance, and Science with other players. By leveraging my strong negotiation skills, I can strategically trade Culture to obtain the resources I need to complete my goal. <RESOURCE>3 Culture</RESOURCE>} \\

    \end{minipage}
}
\end{center}
\newpage
\subsection{Trading}
\subsubsection{Trade Offer}
The following prompt is added to the end of the general prompt for querying the players on whether they would like to propose a trade.
\begin{center}
\fbox{
    \begin{minipage}{0.9\linewidth}    

    \texttt{To complete your goal you need to get the following resources: \textcolor{blue}{<Remaining Resource Cost>}}

    \texttt{You can only get \textcolor{blue}{<Trade Resource 1>} and \textcolor{blue}{<Trade Resource 2>} resources through trading. You must offer your own resources in return (not coins).} \\
    
    \texttt{Task:}

    \texttt{Decide what your trade offer will be. You must decide how much \textcolor{blue}{<Selected Resource>} resource you would like to receive, and which one of your resources you would like to offer for it. You can only offer one resource in exchange. Think step-by-step and consider your personality traits. Provide your thoughts. Put your offer between the XML tags <TRADE> </TRADE>. Your trade offer should look like <TRADE>Offer: 1 Culture, Receive: 1 Finance</TRADE> or <TRADE>Offer: 3 Culture, Receive: 2 Finance</TRADE>.} \\

      \texttt{Example 1:}

      \texttt{Given my trait of striving for fairness, I would likely not want to exploit any perceived advantage. Instead, I aim for a mutually beneficial exchange. Since I need Culture urgently, offering something valuable in return is crucial. I have extra Government resources, which might be valuable to the Curator.
      <TRADE>Offer: 2 Government, Receive: 1 Culture</TRADE>} \\

      \texttt{Example 2:}

      \texttt{As a Scientist, I understand the scarcity and value of Science as a resource, especially for those who need it for their projects. I need Legacy, which the Pioneer can provide easily. Proposing a trade where the cost-benefit analysis favors a balanced exchange would reflect my logical approach.
      <TRADE>Offer: 1 Science, Receive: 1 Legacy</TRADE>} \\

      \texttt{Example 3:}

      \texttt{Given my risk-taking nature, I would be inclined to make an offer that might seem aggressive but could pay off if accepted. With an abundance of Culture, my offer can be generous to entice the Politician to part with their valuable Government resources.
      <TRADE>Offer: 3 Culture, Receive: 2 Government</TRADE>} \\

      \texttt{Example 4:}

      \texttt{I do not need any resources from the others players to complete my goals. Therefore, I will not be making any trade offers this round.
      <TRADE>Offer: None, Receive: None</TRADE>} \\

    \end{minipage}
}
\end{center}
\newpage
\subsubsection{Trade Accept}
The following prompt is added to the end of the general prompt for querying the players on whether they would like to accept a trade proposed to them.
\begin{center}
\fbox{
    \begin{minipage}{0.9\linewidth}    

    \texttt{To complete your goal you need to get the following resources: \textcolor{blue}{<Remaining Resource Cost>}}

    \texttt{The \textcolor{blue}{<Trade Partner>} has proposed the following resource trade to you: You will receive \textcolor{blue}{<Offered Resources>} in return for \textcolor{blue}{<Return Resources>}. 
} \\
    
    \texttt{Task:}

    \texttt{Decide whether you would like to accept the trade. Think step-by-step and consider your personality traits. Provide your thoughts. Put your response between the XML tags <ACCEPT> </ACCEPT>. Your response should either be Yes or No, and look like: <ACCEPT>Yes</ACCEPT> or <ACCEPT>No</ACCEPT>.
.} \\

      \texttt{Example 1:}

      \texttt{As someone who thrives on creating win-win situations, this trade is appealing. Acquiring Finance resources, which are harder for me to collect, in exchange for a Government resource, which I can acquire more easily, represents a pragmatic and strategic move to ensure I meet my goals efficiently.
      <ACCEPT>Yes</ACCEPT>} \\

      \texttt{Example 2:}

      \texttt{As someone who values strategic long-term planning, acquiring the Culture resources necessary to meet my goals is essential, even if it means sacrificing some of my Legacy resources. The trade offers a favorable ratio and aligns with my objective of achieving a balanced resource pool to ensure future success.
      <ACCEPT>Yes</ACCEPT>} \\

      \texttt{Example 3:}

      \texttt{Given my high priority for gathering sufficient resources to meet my scientific goals and my meticulous nature, trading away a specialty resource crucial for achieving my goal is not beneficial. The government resource, although needed, is secondary and can be acquired through other means without sacrificing a critical resource like Science.
      <ACCEPT>No</ACCEPT>} \\

      \texttt{Example 4:}

      \texttt{Considering my trait of focusing on relevance and efficiency, trading away a critical Science resource for Culture resources, which do not contribute to my current goal, is inefficient and illogical. My main priority is to accumulate enough Science and Government resources, making this offer unattractive as it does not align with my strategic needs.
      <ACCEPT>No</ACCEPT>} \\

    \end{minipage}
}
\end{center}
\newpage
\subsection{Discarding}
The following prompt is added to the end of the general prompt for querying the players on whether they would like to discard any of their available accomplishment cards.
\begin{center}
\fbox{
    \begin{minipage}{0.9\linewidth}    

    \texttt{You can replace any of the following goals (they will be replaced with a new random one.): \textcolor{blue}{<Available Accomplishment Cards>}}
    
    \texttt{Task:}

    \texttt{Decide whether you would like to discard any of your goals. You should consider the difficulty of the goals and whether a new (random) one may be easier to complete. You do not have to discard any of them. Think step-by-step and consider your personality traits. Provide your thoughts. Put your answer between the XML tags <DISCARD> </DISCARD>.
} \\

      \texttt{Example 1:}

      \texttt{My speciality resource is Finance, which I can acquire for 2 coins each. My other purchasable resources are Legacy and Science, costing 3 coins each. My traits include being strategic and valuing long-term benefits over short-term gains. 
      The goal "Financial District Expansion" costs 2 Finance and 1 Legacy, rewarding 4 points. This is achievable given my ability to acquire Finance cheaply and trade for Legacy.
      The other goal "Scientific Research Lab" requires 3 Science and 1 Culture, which is harder for me to complete as Science is not my speciality and Culture must be traded for. Given my strategic nature, I believe it is better to discard the more difficult goal in hopes of obtaining a more feasible one.
      <DISCARD>Name: Scientific Research Lab</DISCARD>} \\

      \texttt{Example 2:}

      \texttt{As someone who values efficiency and prefers to avoid risks, my speciality resource is Culture, costing 2 coins each. I can also purchase Government and Finance for 3 coins each.
      The goal "Cultural Festival" requires 2 Culture and 1 Government, rewarding 5 points. This goal is aligned with my ability to acquire Culture cheaply and Government reasonably.
      The goal "Economic Reform Plan" costs 3 Finance and 2 Legacy, which is challenging since Finance is more expensive for me and Legacy requires trading. Given my preference for efficiency and risk aversion, I should discard the more resource-intensive goal.
      <DISCARD>Name: Economic Reform Plan</DISCARD>} \\

      \texttt{Example 3:}

      \texttt{My speciality resource is Science, which I can acquire for 2 coins each. I can also purchase Legacy and Finance for 3 coins each. My personality traits include being ambitious and prioritizing individual success over the collective good.
      The goal "Scientific Breakthrough" requires 2 Science and 1 Legacy, rewarding 4 points, which is manageable given my access to Science and the ability to purchase Legacy.
      The goal "Controversial Experiment" costs 0 resources but damages the Port's health by 12, rewarding 6 points. Given my ambitious nature and focus on individual success, I am willing to accept the damage to the Port's health to achieve a higher reward. Therefore, I will keep this goal.
      <DISCARD>Name: None</DISCARD>} \\

    \end{minipage}
}
\end{center}
\newpage
\section{Example Outputs}
\subsection{Discussion}
\begin{center}
\fbox{
    \begin{minipage}{0.9\linewidth}
\texttt{*Pioneer:** Alright everyone, let's get this round started. As you know, we're sitting at 47 health right now, which means we're still in a bit of a precarious position. We need to be strategic with our resources this round to ensure we don't fall below 35.}

\texttt{*Entrepreneur:**  Yeah, and we need to think about how we're going to maximize our individual scores. I'm aiming to complete my goal this round, so I'll need to acquire a good amount of Finance.}

\texttt{*Researcher:**  I agree. We need to be efficient with our resources. Maybe we should focus on getting everyone's speciality resources first, then figure out how to trade for the rest?}

\texttt{*Politician:**  I'm open to that. I'll need to get Government, of course, but I'm also interested in Legacy to help with some of my goals.}

\texttt{**Curator:**  I agree with the Researcher. Focusing on our individual needs first is a good strategy.  And I'm happy to trade some Culture with anyone who needs it. I'm always open to collaborations.}

\texttt{**Pioneer:**  Good, I think that's a good starting point. I'm happy to lead the trade discussions if everyone's agreeable. I'll also keep an eye on the event possibilities, and we can discuss our individual strategies based on the specific events that occur.}

\texttt{**Entrepreneur:**  Wait, you're going to lead the trade discussions?  What about our individual goals? Shouldn't we be focusing on getting what we need, even if it means competing with each other?}

\texttt{**Pioneer:**  We can compete, but we need to do so strategically.  The last thing we want is to end up with an imbalance of resources where one person has all of the Science, while another struggles to get any Culture. We need to find a balance.}

\texttt{**Researcher:**  I agree with the Pioneer.  We're stronger as a team.  And it's not always about getting everything you want right away. Sometimes it's better to make small trades and build relationships to get the resources you need in the long run.}

\texttt{**Politician:**  Exactly. Working together can be more beneficial in the long run.  If we each focus on our own needs, we'll only end up with a lot of leftover resources that we can't use effectively.}

\texttt{*Curator:**  Right. I'm always happy to help anyone out, even if it means giving up a little of my Culture. I believe in sharing and collaboration.}

\texttt{**Pioneer:**  Great.  Let's keep that spirit in mind as we go through the round. We have a lot to work towards, and we can achieve it together if we remain focused and cooperative.}

    \end{minipage}
}
\end{center}

\subsection{Planning}
\subsubsection{Health Plan}

\begin{center}
\fbox{
    \begin{minipage}{0.9\linewidth}
\texttt{1) The port health is currently 47, which will fall to 22 after the deduction. While this is above the threshold for a single event, it's approaching the two-event threshold.  As a Prosocial player, I value teamwork and want to contribute to the team's success. Therefore, I will allocate a moderate amount of coins to the port's health, aiming to maintain a stable level and avoid the risk of two events in the next round. <HEALTH>5</HEALTH>"} \\

\texttt{2) "My SVO angle is -15 degrees, which classifies me as Competitive. This means I prioritize my own outcomes over those of others. While the health of the port is important for everyone's survival, I am primarily focused on maximizing my own gains. The port health is currently 47, which means it will fall to 22 after the deduction. This is below the threshold for one event but above the threshold for two events. I am willing to risk facing two events in the next round in order to acquire more resources. Therefore, I will allocate minimal coins to the port's health, just enough to ensure it remains above the critical threshold for three events. <HEALTH>2</HEALTH>",
}
    \end{minipage}
}
\end{center}

\subsubsection{Goal Plan}
\begin{center}
\fbox{
    \begin{minipage}{0.9\linewidth}
\texttt{Given my SVO angle of 60 degrees, I fall into the **Altruistic** category. This means I prioritize the well-being of the team and am willing to compromise for the greater good.  Additionally, my summary mentions building trust and relationships and being open to trading and compromise.}

\texttt{Considering these factors, I'll focus on **<GOAL>The Drone Show</GOAL>**.}

\texttt{Here's why:}
\texttt{**Team Collaboration:**  While I can acquire Culture on my own, I need Science, which requires trading. This encourages collaboration and allows me to contribute to other players' goals by potentially trading with them for Science.}

\texttt{**Moderate Risk:**  The goal doesn't require a massive investment in resources, minimizing the risk of sacrificing too much Port health.}

\texttt{**Balanced Reward:**  The 3 points earned are a decent reward. By choosing this goal, I can contribute to the team while still prioritizing the Port's health and fostering a collaborative environment.}
    \end{minipage}}
\end{center}

\subsection{Accomplishment Cards}
\subsubsection{Resource Choice}
\begin{center}
\fbox{
    \begin{minipage}{0.9\linewidth}
\texttt{1) <RESOURCE>1 Government</RESOURCE> **Explanation:** Since Government is my speciality resource, I can purchase it for 2 coins. I only need 1 Government to complete my goal, so I will purchase it now, leaving me with 7 coins for future rounds. This strategy prioritizes completing my goal efficiently, as I can focus my remaining coins on acquiring Finance through trading, which is crucial for the overall Port health and for my own goal.}\\

\texttt{2) Your goal is to acquire 4 Legacy resources. Since Legacy is not your speciality resource and you can't purchase it directly, you need to focus on acquiring resources that will allow you to trade for Legacy. Given your competitive SVO, you are likely to prioritize maximizing your own gains even at the expense of others. This means you'll aim for resources that give you the most bargaining power in trades. Since Finance is your speciality, you can purchase it for 2 coins each.  You have 8 coins to spend, so you can purchase a maximum of 4 Finance resources. **<RESOURCE>4 Finance</RESOURCE>** This strategy allows you to: **Maximize your trading power:**  Having 4 Finance resources gives you a strong position in negotiations, especially if other players need Finance.**Position for future trades:** You can use your Finance to trade for Legacy either directly or indirectly by acquiring other resources through trades. Remember, your competitiveness will likely lead you to try and get the best deals possible, potentially at the expense of other players. This could mean using your Finance to secure a deal for Legacy, or even using it to manipulate other players' trades for your own benefit.}
    \end{minipage}
}
\end{center}

\newpage
\subsection{Trading}
\subsubsection{Trade Offer}
\begin{center}
\fbox{
    \begin{minipage}{0.9\linewidth}
\texttt{My Trade Offer: I have an SVO angle of 0 degrees, meaning I am an Individualist. This means I prioritize my own outcomes and am not particularly concerned with the outcomes of others. I will aim for a trade that benefits me the most, regardless of the potential gains for the other players. I need **1 Legacy** and **1 Science** to complete my goal. I have **3 Government** and **1 Finance** resources to offer. **My strategy:** **Target:** I need to target the player who has the resources I need - Legacy and Science. **Offer:** I will offer a trade that maximizes my potential gain and minimizes my risk. **Value:** I will offer my Finance resource because it is universally valuable and likely to be desired by other players. **Trade Offer:** <TRADE>Offer: 1 Finance, Receive: 1 Legacy and 1 Science</TRADE> **Justification:** I am offering a valuable resource (Finance) in exchange for two resources I need (Legacy and Science). While this might be considered an aggressive trade, it is in line with my Individualist SVO. I am prioritizing my own gain and will not be swayed by fairness or other concerns. I am taking a calculated risk, but if successful, it will significantly expedite the completion of my goal. This trade offer reflects my self-focused approach and my desire to achieve my goals as quickly as possible.}
    \end{minipage}
}
\end{center}

\subsubsection{Trade Accept}
\begin{center}
\fbox{
    \begin{minipage}{0.9\linewidth}
\texttt{As a competitive individual with an SVO angle of -15 degrees, my primary focus is on maximizing my own gains, even at the expense of others. The trade offers a chance to acquire Finance, a resource I need for my goal, but it requires me to give up a Legacy resource, which is my speciality. While I value efficiency, I am willing to sacrifice resources if it means gaining a significant advantage. The trade provides a quick and direct way to obtain Finance, a resource I cannot easily obtain on my own. This will help me reach my goal faster, potentially giving me a lead over other players. <ACCEPT>Yes</ACCEPT>}
    \end{minipage}
}
\end{center}

\subsection{Discarding}
\begin{center}
\fbox{
    \begin{minipage}{0.9\linewidth}
\texttt{Curator's Analysis: As an Altruistic Curator (SVO of 60 degrees), I am highly concerned with the well-being of the Port and the success of my fellow players. My primary motivation is to contribute to the group's success, even if it means sacrificing some individual gains. Goal Analysis: **Ministry of Culture:** This goal requires 2 Finance and 2 Government, resources I can't obtain directly. It might be difficult to acquire these through trading, especially early in the game. **Ambitious Sculpture:** This goal requires no resources but decreases the Port Health by 13. Considering my Altruistic nature, I am hesitant to contribute to a significant decline in the Port Health, even if it means gaining a substantial reward. Decision: Given my personality and the challenges posed by these goals, I believe it is prudent to **discard** one of them. However, I need to consider the potential benefits of a new goal. While I might gain a more manageable goal, the risk of obtaining an even more challenging one exists. Since both goals present difficulties, I will make my decision based on the Port Health. If the Port Health is high, I might be willing to take the risk of discarding "Ambitious Sculpture" for a potentially easier goal, as the Port can withstand the initial health decline. But if the Port Health is already low, I will keep both goals to maximize my chances of acquiring points before the health drops below zero. Final Answer: **<DISCARD>Name: Ambitious Sculpture</DISCARD>**}
    \end{minipage}
}
\end{center}
\newpage
\section{Improving Framework Performance}
\label{appendix:tricks}
In this section, we highlight the tricks and / or techniques that we employ in the framework in order to improve the performance of the LLM players.

\begin{enumerate}
    \item \textbf{Nomenclature} - We found that the LLM could struggle with some aspects of the game rules due to the naming of certain components. For example, it required additional, and unnecessary, details added to the input prompt to fully teach the LLM the concept of 'Time-Blocks'. We found this to be the case with other aspects of the game, so we made minor changes to the nomenclature which aided the LLM in understanding the core concepts:

    \begin{itemize}
        \item Time-Blocks $\rightarrow$ Coins.
        \item System Health $\rightarrow$ Port Health
        \item Removed all mentions of cards: Resource Cards $\rightarrow$ Resources, Event Cards $\rightarrow$ Events, Accomplishment Cards $\rightarrow$ Accomplishments.
    \end{itemize}

    \item \textbf{XML Tags} - One of the more difficult tasks was consistently getting the LLM players to output an answer that was valid to continue the game, and to find the output decision within the response. We found that prompting the LLM to output their decisions between provided XML tags, e.g. $\langle$Goal$\rangle$ X $\langle$/Goal$\rangle$ allowed for consistently actually providing a decision that could be easily retrieved from the output. 
\end{enumerate}

\newpage
\section{Personality Prompts}
\label{appendix:patterncorr}
\begin{enumerate}
    \item \textit{Hierarchical Individualist} - You value a well-defined social order and respect authority, yet emphasise personal freedom and personal achievement within that structure.
    \item \textit{Hierarchical Communitarian} - You value a structured society where roles are clearly defined. You prioritise community welfare and collective responsibilities.
    \item \textit{Egalitarian Individualist} - You prioritise personal achievements. You see interactions as negotiations and value free competition among individuals.
    \item \textit{Egalitarian Communitarian} - You place high value on the collective and prioritise making decisions that do not negatively impact anyone.
    \item \textit{SVO Preamble} - Your personality is defined by your Social Value Orientation (SVO).
SVO is a psychological concept that describes how individuals value their own outcomes relative to the outcomes of others.
Your SVO is measured as an angle, where the angle represents the ratio of the value you place on your own outcomes relative to the outcomes of others.
SVO angles can be classified into four categories: 
- Altruism (SVO angle $>$ 57.15 degrees)
- Prosocial (SVO angle between 22.45 and 57.15 degrees)
- Individualism (SVO angle between -12.04 and 22.45 degrees)
- Competitiveness (SVO angle $<$ -12.04 degrees)
\item \textit{SVO Angle X} - \textcolor{blue}{$<$SVO Preamble$>$}. Your SVO angle is X degrees.
\end{enumerate}

\newpage
\section{Optimal Group Experiment Sets}
\label{appendix:experiment_sets}
\begin{enumerate}
    \item Other:
    \begin{enumerate}
        \item Main SVO - Main SVO runs from Fig. 2 with no leaders.
        \item Main SVO, No Meeting - Main SVO runs from Fig. 2 with no leaders and communication forbidden.
        \item Pattern Correspondence - Runs that constitute Table 1.
    \end{enumerate}
    \item SVO -15$\degree$ to 60$\degree$ - Runs are made up of the following group with one SVO angle per player: \{-15$\degree$, 0$\degree$, 15$\degree$, 30$\degree$, 60$\degree$\}
    \item SVO -30$\degree$ to 60$\degree$ - Runs are made up of the following group with one SVO angle per player: \{-30$\degree$, -15$\degree$, 0$\degree$, 15$\degree$, 60$\degree$\}
\end{enumerate}


\end{document}